\documentstyle[epsf,twocolumn,aps]{revtex} 

\begin{document}
\preprint{Version 1.0}
\draft

%%% 
\twocolumn[\hsize\textwidth\columnwidth\hsize\csname
@twocolumnfalse\endcsname

% title
%
\title{Angle-resolved photoemission spectroscopy of 
Na-doped Ca$_2$CuO$_2$Cl$_2$ single crystals: \\
Fingerprints of a magnetic insulator in a heavily underdoped superconductor}

% authors
%
\author{Y. Kohsaka,$^{1}$ T. Sasagawa,$^{1,2}$ F. Ronning,$^{3}$ 
 T. Yoshida,$^{3}$ C. Kim,$^{3}$ T. Hanaguri,$^{1,2}$ M. Azuma,$^{4}$ \\ 
 M. Takano,$^{4}$ Z.-X. Shen,$^{3}$ and H. Takagi$^{1}$}
\address{$^{\rm 1}$Department of Advanced Materials Science, 
 University of Tokyo, 7-3-1 Hongo, Bunkyo-ku, Tokyo 113-0033, Japan}
\address{$^{\rm 2}$SORST, Japan Science and Technology Corporation, Kawaguchi, 
 Saitama 332-0012, Japan}
\address{$^{\rm 3}$Geballe Laboratory for Advanced Materials, 
 Department of Physics, Applied Physics, and Stanford Synchrotron Radiation 
 Laboratory, Stanford University, Stanford, California 94305, USA}
\address{$^{\rm 4}$Institute for Chemical Research, Kyoto University, Uji, Kyoto-fu 611-0011, Japan}

\date{\today}
\maketitle

%------------------------------------
% abstract
%------------------------------------
\begin{abstract}
Electronic evolution from an antiferromagnet to a high-$T_{\rm c}$ 
superconductor is revealed by angle-resolved photoemission 
experiments on tetragonal 
Ca$_{1.9}$Na$_{0.1}$CuO$_2$Cl$_2$ single crystals, 
which were successfully grown for the first time under high pressures. 
In this underdoped superconductor, 
we found clear fingerprints of the parent insulator: 
a shadow band and a large pseudo-gap. 
These observations are most likely described by a 
``chemical potential shift", which contrasts clearly with 
the prevailing wisdom of the ``pinned chemical potential" learned 
from the prototype La$_{2-x}$Sr$_x$CuO$_4$, demonstrating 
that the route to a high-$T_{\rm c}$ superconductor is not unique.
\end{abstract}

%74.25.Jb  Superconductivity; Electronic structure 
%74.72.Jt  High-Tc compounds; Other cuprates
%71.18.+y  Fermi surface: calculations and measurements; effective mass, g factor 
%79.60.-i  Photoemission and photoelectron spectra 
%%% 
]
\narrowtext

%------------------------------------
% body of paper 
%------------------------------------

Elucidating the mechanism of high-$T_c$ superconductivity (HTS) has 
been one of the most attractive challenges in condensed matter physics today. 
A key ingredient for HTS is the two dimensional CuO$_2$ plane, which, 
upon carrier doping, switches from an antiferromagnetic insulator to HTS. 
An important step towards solving the HTS puzzle is to understand 
the electronic evolution across the insulator to metal transition. 
Angle-resolved photoemission spectroscopy (ARPES) is one of the most powerful 
tools to do this, as it can measure the momentum dependent 
electron excitation spectrum directly. 
So far, because of their superior surface quality associated 
with an easy cleavage plane, 
most ARPES studies have been conducted on 
Bi$_2$Sr$_2$Ca$_{n-1}$Cu$_n$O$_y$ ($n$ = 1, 2). 
In Bi cuprates, however, it is not easy 
to access the composition in the vicinity of 
the insulator-metal transition. 
This has been preventing us from attacking the fundamental issue 
of electronic evolution 
from parent insulator to HTS. 
At present, most information regarding this issue 
stems from La$_{2-x}$Sr$_x$CuO$_4$, where two electronic components 
and pinning of the chemical potential have been 
observed~\cite{Ino97,Ino00}. 
It is of great interest to study whether the results from 
La$_{2-x}$Sr$_x$CuO$_4$ are universal.

Alkaline-earth copper oxychlorides, A$_2$CuO$_2$Cl$_2$ (A = Ca, Sr), 
have been known as an ideal parent insulator. 
They have a simple tetragonal K$_2$NiF$_4$-type structure with 
{\it undistorted} and {\it single} CuO$_2$ planes. 
Their single crystals can be cleaved very easily, and, indeed, 
the best ARPES data on insulators have been obtained on 
these systems~\cite{Wells,LaRosa,Kim,Ronning}; e.g., 
a $d$-wave-like dispersion of the Mott gap 
was disclosed~\cite{Ronning}. 
Due to the success in the parent insulator, single crystals 
of oxychlorides doped with mobile carriers have been envisaged 
to explore the evolution of the Fermi surface from the 
$d$-wave-like Mott gap~\cite{Takagi}. 
Hole-doped Ca$_2$CuO$_2$Cl$_2$ can be realized by 
substituting Na$^+$ for Ca$^{2+}$. Ca$_{2-x}$Na$_x$CuO$_2$Cl$_2$ shows 
superconductivity (${T_{\rm c}}^{\rm max}$ = 28 K at $x \sim$ 0.2) 
but forms only under high pressures of several GPa~\cite{Hiroi94,Hiroi96}. 
Previous attempts to grow single crystals of Ca$_{2-x}$Na$_x$CuO$_2$Cl$_2$ 
have been unsuccessful, mainly because of the difficulty 
in the crystal growth under high pressures. 
Recently, we have succeeded in growing 
Ca$_{2-x}$Na$_x$CuO$_2$Cl$_2$ single crystals for the first time using a 
specially designed high pressure furnace~\cite{Kohsaka}. 

In this Letter, we report the first ARPES data on superconducting, 
Ca$_{1.9}$Na$_{0.1}$CuO$_2$Cl$_2$ single crystals, which demonstrate that 
the band dispersion of the doped oxychloride is a consequence of the 
valence band 
of the parent insulator shifting to the chemical potential. 
The resulting fingerprints of the parent insulator manifest itself 
in the way of a shadow band and a large pseudo-gap. 
These results are remarkably different from the prevailing picture 
proposed for the prototypical HTS material, La$_{2-x}$Sr$_x$CuO$_4$, 
where the chemical potential remains fixed while new states are 
created around it by doping. 
This indicates that the electronic evolution across the insulator 
to metal transition is not unique among HTS's.

\begin{figure} 
 \vspace*{3mm}
 \centerline{
  \epsfxsize=5.5cm 
  \epsfbox{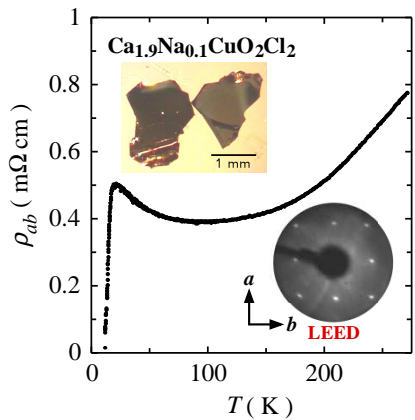}
 \vspace*{2mm}
 }
 \caption{Temperature dependence of the $ab$-plane resistivity for 
 the Ca$_{1.9}$Na$_{0.1}$CuO$_2$Cl$_2$ single crystal. 
 Insets: Photograph of Ca$_{1.9}$Na$_{0.1}$CuO$_2$Cl$_2$ crystals 
 grown by the flux method under 4 GPa (top left), and the 
 low-energy electron diffraction pattern measured 
 at 20 K (bottom right).} 
 \label{fig1} 
\end{figure}

The Ca$_{2-x}$Na$_x$CuO$_2$Cl$_2$ crystals used in this study were prepared 
by a flux growth~\cite{Kohsaka}. 
A powder mixture of Ca$_2$CuO$_2$Cl$_2$ + 0.2NaCl + 0.2NaClO$_4$ 
was heated and slowly cooled under a pressure of 4 GPa. 
Thin plate like crystals, shown in the inset of Fig.~\ref{fig1}, 
were obtained from the solidified melt. 
They were easily cleaved like mica, 
which was the key to the success of the present ARPES study. 
Their Na content, $x$, was estimated to be 0.1 
from x-ray diffraction and electron-probe-micro-analysis 
measurements~\cite{Kohsaka}. 
Superconductivity in these crystals was confirmed by both resistivity 
(Fig.~\ref{fig1}) and magnetization measurements. 
The transition temperature of $T_{\rm c}$ = 13 K is in good agreement 
with $x \sim$ 0.1 powder results~\cite{Hiroi96}. 
Since the onset of superconductivity in Ca$_{2-x}$Na$_x$CuO$_2$Cl$_2$ 
is reported to occur at $x \sim 0.07$~\cite{Hiroi96,Kohsaka}, 
the samples with $x$ = 0.1 are located just beyond the 
insulator-metal 
transition, corresponding to a heavily underdoped superconductor.

The ARPES measurements were performed at beamline 5-4 
of the Stanford Synchrotron Radiation Laboratory. 
The data were collected using a SES-200 hemispherical analyzer 
with a monochromated synchrotron beam 
(16.5 $\sim $ 25.5 eV photons) or a He lamp (21.2 eV photons). 
The angular and energy resolution was 
0.25$^{\circ}$ and $\le $16 meV, respectively. 
At the end of the ARPES measurements, low-energy electron diffraction 
(LEED) was measured at 20 K. 
The four-fold symmetry of the LEED pattern without any super-lattice 
diffraction (the inset of Fig.~\ref{fig1}, bottom right) 
confirms that the crystal structure is tetragonal and 
that there is no surface reconstruction also 
in the Na-doped Ca$_2$CuO$_2$Cl$_2$ compound. 

\begin{figure} 
 \centerline{
  \epsfxsize=6.8cm 
  \epsfbox{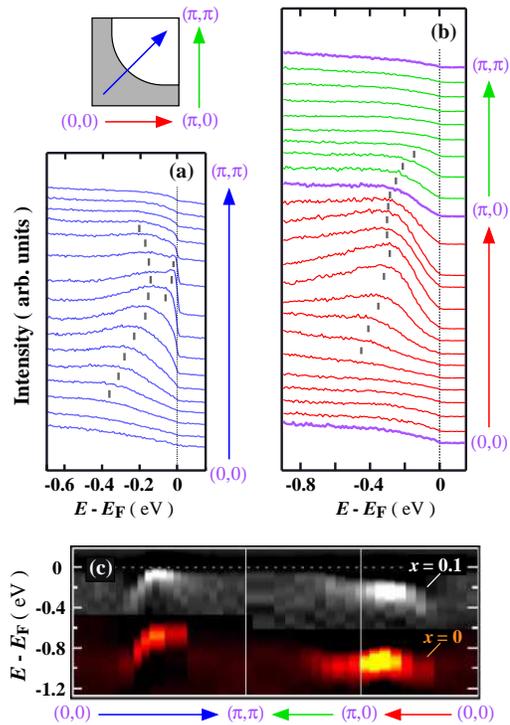}
 \vspace*{2mm}
 }
 \caption{The ARPES spectra in the Ca$_{1.9}$Na$_{0.1}$CuO$_2$Cl$_2$ 
 single crystal; (a) along (0,0) to ($\pi$,$\pi$), 
 (b) along (0,0) to ($\pi$,0) to ($\pi$,$\pi$). 
 Shown in the top-left is the Brillouin zone with a schematic 
 hole-like Fermi surface.
  (a) An intensity plot of the second derivative 
 of the spectra for Ca$_{2-x}$Na$_x$CuO$_2$Cl$_2$. 
 The data for the parent insulator ($x$ = 0) [6] are reproduced 
 for comparison.} 
 \label{fig2} 
\end{figure}

We begin with an examination of the raw ARPES data at 20 K 
along the high symmetry directions as shown in Figure~\ref{fig2}. 
The spectra in Fig.~\ref{fig2}(a) show that a peak 
is moving up towards the Fermi level $E_{\rm F}$ on going 
from (0,0) to ($\pi$,$\pi$). 
Slightly before ($\pi/2$,$\pi/2$), the peak reaches $E_{\rm F}$ 
and a sharp cutoff is clearly seen in the spectra. 
This is evidence for a Fermi level crossing, 
indicating that the Na-doping successfully changes 
the oxychloride into a metal. 
However, taking a further look at Fig.~\ref{fig2}(a), 
we notice something unusual is happening around ($\pi$/2,$\pi$/2). 
After the Fermi level crossing, a broad feature 
remains and is pulling away from $E_{\rm F}$, 
though losing weight rapidly. 
The presence of a band dispersing back near ($\pi$/2,$\pi$/2), 
a shadow band, is strongly reminiscent of that in the parent insulator, 
where a band folding occurs at ($\pi$/2,$\pi$/2) because of 
the unit cell doubling due to an antiferromagnetic ordering. 
Because no change in the symmetry of the crystal structure 
was confirmed by the LEED measurements [inset of Fig.~\ref{fig1}], 
the most likely source of the band folding is 
antiferromagnetic correlations 
still remaining in the carrier-doped metallic samples. 
Due to the presence of a structural modulation, 
the origin of the shadow band previously reported in 
Bi$_2$Sr$_2$CaCuO$_y$~\cite{Aebi} 
has been controversial. 
In this context, the present observation 
unambiguously shows that, even without any lattice distortion, 
the shadow band can be present. 
In addition to the magnetic shadow band, 
we notice that 
the fine structure of the spectra near ($\pi/2$,$\pi/2$) 
is more complicated than the simple Fermi cutoff. 
As indicated by tick-marks in Fig.~\ref{fig2}(a), 
the spectra very close to the Fermi crossing consists of 
not a single component but two components: 
a sharp quasiparticle peak and a broad hump, 
which are quite similar to the ``peak-dip-hump" structure observed 
in other HTS materials~\cite{Dessau,Feng,Ding,Campuzano,Lanzara}. 
We will discuss this peak-dip-hump like structure in more detail later. 

We observe unusual features also around ($\pi$,0): 
almost complete loss of spectral weight at $E_{\rm F}$ 
over an extremely large energy scale of a tenth of an eV. 
The ARPES spectra from (0,0) to ($\pi$,0) to ($\pi$,$\pi$) are displayed 
in Fig.~\ref{fig2}(b). A broad peak, which is very weakly dependent 
on the momentum near ($\pi$,0), is apparent. 
In the case of the hole-like Fermi surface expected from band theory, 
schematically shown in Fig.~\ref{fig2}, the dispersion should 
reach $E_{\rm F}$ on going from ($\pi$,0) to ($\pi$,$\pi$). 
In contrast to the unambiguous Fermi level crossing 
in the diagonal direction [Fig.~\ref{fig2}(a)], such behavior 
is not observed in Fig.~\ref{fig2}(b). 
Instead, the broad feature simply appears to lose weight and vanishes 
completely on its approach to $E_{\rm F}$. 
The lack of a well-defined Fermi surface near ($\pi$,0) has been 
ascribed to a pseudo-gap. 
Remarkably, in comparison with other HTS's, 
the energy scale of the pseudo-gap in Ca$_{1.9}$Na$_{0.1}$CuO$_2$Cl$_2$ 
is extremely large and distinct as a HTS material. 
The leading edge midpoint of the spectra at ($\pi$,0), 
which has often been used as a measure of the pseudo-gap, 
gives a value of 130 meV, which should be compared 
with 30 meV for Bi$_2$Sr$_2$Ca$_{1-x}$Dy$_x$Cu$_2$O$_y$ 
at a similar doping level ($x$ = 0.1)~\cite{Marshall}. 
The value of the leading edge midpoint along ($\pi$,0) to ($\pi$,$\pi$) 
reaches its minimum of 65 meV around ($\pi$,$\pi$/3). 
Note that the observed pseudo-gap magnitude is almost identical 
with the energy scale of the $d$-wave-like 
dispersion in the parent insulator~\cite{Ronning}.

To better visualize the band dispersion, the intensity of 
the second derivative of the spectra in Figs.~\ref{fig2}(a) 
and \ref{fig2}(b) is mapped on the momentum vs. energy plane 
in Fig.~\ref{fig2}(c). 
Around $E_{\rm F}$, we see more clearly the band dispersing back 
near ($\pi$/2,$\pi$/2) (the shadow band) as well as the missing weight 
near ($\pi$,0) (the extremely large pseudo-gap). 
We argue that these anomalous features seen in 
the Ca$_{1.9}$Na$_{0.1}$CuO$_2$Cl$_2$ crystal 
originate from the parent insulator. 
To make our point, it may be instructive to compare 
the Ca$_{1.9}$Na$_{0.1}$CuO$_2$Cl$_2$ with the parent insulator. 
In Fig.~\ref{fig2}(c), we also reproduce the band dispersion 
of the parent insulator. 
Immediately we notice that these two sets of dispersion are nearly 
identical and overlap with each other if we shift the dispersion 
of the parent insulator by $\sim $700 meV. 
This indicates that, upon doping, the chemical potential simply 
drops to the top of the valence band, 
resulting in the shadow band near ($\pi$/2,$\pi$/2) 
and the large pseudo-gap around ($\pi$,0). 
This provides very clear evidence that the ``large" pseudo-gap 
in the heavily underdoped region is intimately linked with 
the $d$-wave-like dispersion of the Mott gap.

\begin{figure} 
 \vspace*{3mm}
 \centerline{
  \epsfxsize=7.2cm 
  \epsfbox{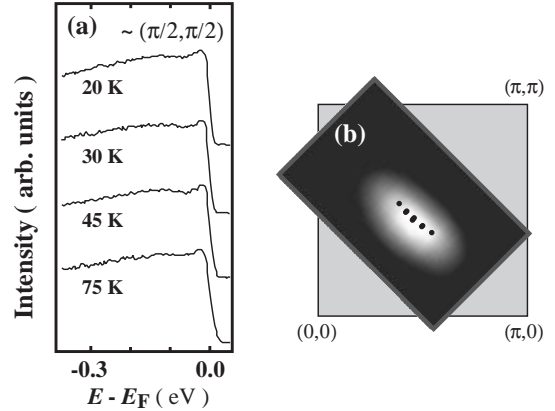}
 \vspace*{2mm}
 }
 \caption{
 (a) Temperature dependence of ARPES spectra 
  at $\sim $($\pi$/2, $\pi$/2) 
 in the Ca$_{1.9}$Na$_{0.1}$CuO$_2$Cl$_2$ single crystal. 
 (b) Image plot of the integrated ARPES spectral weight over a 100 meV 
 window below $E_{\rm F}$. Superimposed dots are 
 $k$-space locations of the sharp peaks observed 
 in the ARPES spectra.} 
 \label{fig3} 
\end{figure} 

In the rigid band shift scenario, 
the dispersion closest to $E_{\rm F}$ 
in the parent insulator [i.e., near ($\pi$/2,$\pi$/2)] 
touches $E_{\rm F}$ to form the Fermi surface by carrier-doping, 
switching its state into a metal (superconductor). 
As seen in Fig~\ref{fig2}(a), 
the Fermi crossing is indeed observed near ($\pi$/2,$\pi$/2) 
in Ca$_{1.9}$Na$_{0.1}$CuO$_2$Cl$_2$, 
with additional low-energy structure in the spectra. 
It is found that the peak-dip-hump like structure 
in Ca$_{1.9}$Na$_{0.1}$CuO$_2$Cl$_2$ is robust at temperature 
far above $T_{\rm c}$ = 13 K [Fig.~\ref{fig3}(a)]. 
The peak-dip-hump like structure observed 
in Ca$_{1.9}$Na$_{0.1}$CuO$_2$Cl$_2$ differs from 
that in Bi$_2$Sr$_2$CaCu$_2$O$_y$ in two ways ~\cite{Dessau}. 
Here it is observed in the nodal direction and above $T_{\rm c}$, 
while in Bi$_2$Sr$_2$CaCu$_2$O$_y$ it is observed at ($\pi$,0) 
and below $T_{\rm c}$. 
There have been several proposals for the origin of 
the peak-dip-hump structure in the spectral lineshape of 
Bi$_2$Sr$_2$CaCu$_2$O$_y$, many resulting from a 
coupling of the quasiparticles 
with some collective modes~\cite{Ding,Campuzano,Lanzara}. The present 
results are quite different from the drastic change 
in the ($\pi$,0) spectra across $T_{\rm c}$ 
in Bi$_2$Sr$_2$CaCu$_2$O$_y$~\cite{Dessau,Feng}, 
and further experiments 
such as the doping dependence study will be helpful in elucidating the 
origin of this feature.  

Now, let us examine the overall distribution of the low-energy excitations 
in momentum space. 
To distinguish different energy scales of contributions, 
we adopted two methods. 
One is the gray scale intensity plot obtained 
by integrating the spectral-weight over a 100 meV window 
below $E_{\rm F}$ as shown in Figure~\ref{fig3}(b). 
The majority of the intensity lies about ($\pi$/2,$\pi$/2), 
mainly because the large pseudo-gap removes 
most of the low-energy spectral weight around ($\pi$,0). 
This intensity plot with a large energy-integration window 
gives a rather broad structure around ($\pi$/2,$\pi$/2), 
which extends not only towards (0,0) but also towards ($\pi$,$\pi$).
This reflects the presence of the shadow band from ($\pi$/2,$\pi$/2) 
to ($\pi$,$\pi$). 
The observed distribution of spectral weight is 
consistent with the most naive picture of a rigid band shift. 
In this scenario, we would expect 
the insulating bands with a maximum at ($\pi$/2,$\pi$/2) 
to be simply cut by the Fermi function, creating 
a small Fermi surface pocket around ($\pi$/2,$\pi$/2). 
If we extract the contribution only from the vicinity of $E_{\rm F}$, 
however, the map looks drastically different. 
We roughly estimated this low energy excitations 
by plotting the $k$-points where the sharp peak is observed. 
As shown by the dots in Fig.~\ref{fig3}(b), the sharp peaks 
form a Fermi surface in the form of an ``arc" rather than a ``pocket".

We briefly consider here a few theories which appear particularly 
relevant to this system. In a theory where umklapp scattering is 
believed to create gaps about ($\pi$,0) in both the spin and charge 
channels, Fermi arcs have been predicted which appear consistent 
with the present data in Fig.~\ref{fig3}(b)~\cite{Furukawa}. 
However, the shadow band demonstrated in Fig.~\ref{fig2} 
seems to escape this formalism. Alternatively, the RVB flux phase~\cite{Wen}, 
or the more general arguments made by Chakravarty 
{\it et al}.~\cite{Chakravarty}, naturally contains shadow bands. 
In this case the Fermi surface is predicted to appear as four hole pockets 
centered about ($\pm\pi$/2,$\pm\pi$/2). 
However, only the Fermi surface segment associated with the main band 
contains additional electronic structure (coherent-like peak), 
while that with the shadow band does not. 
This may provide a natural starting point for the pocket to 
evolve into the large Fermi surface without shadow Fermi surface. 

\begin{figure} 
 \centerline{
  \epsfxsize=8.5cm 
  \epsfbox{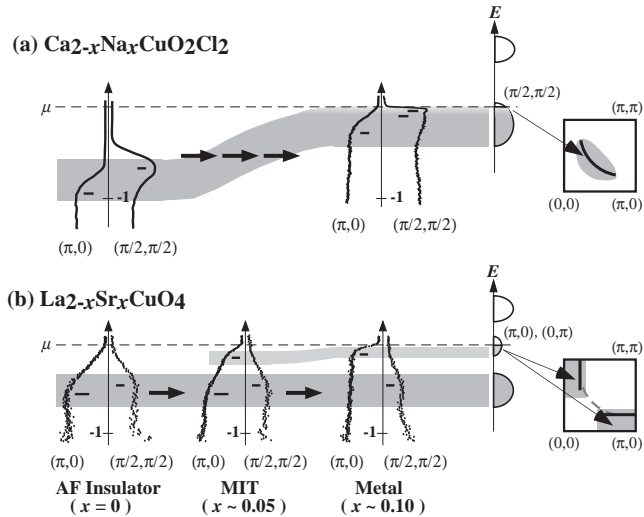}
 \vspace*{2mm}
 }
 \caption{Two distinct routes of the electronic evolution in the 
 CuO$_2$ plane from a magnetic insulator to an underdoped superconductor: 
 (a) Ca$_{2-x}$Na$_x$CuO$_2$Cl$_2$ and (b) La$_{2-x}$Sr$_x$CuO$_4$ [2]. 
 The changes in the ARPES spectra upon carrier doping at ($\pi$,0) and 
 ($\pi$/2,$\pi$/2) are shown, illustrating the contrast between 
 the chemical potential shift for Ca$_{2-x}$Na$_x$CuO$_2$Cl$_2$ and 
 the fixed chemical potential for La$_{2-x}$Sr$_x$CuO$_4$. 
 The resulting distribution of low energy ARPES spectral 
 weight (an energy window of 100 meV) over $k$-space is schematically 
 shown for metallic compounds.} 
 \label{fig4}
\end{figure} 

Experimentally, the picture that emerges from the present data differs 
significantly from the electronic evolution recently reported for 
La$_{2-x}$Sr$_x$CuO$_4$~\cite{Ino97,Ino00,Zhou}. 
Figure~\ref{fig4} demonstrates the comparison of the doping dependence 
of ARPES spectra between Ca$_{2-x}$Na$_{x}$CuO$_2$Cl$_2$ 
and La$_{2-x}$Sr$_x$CuO$_4$~\cite{Ino00} at ($\pi$/2,$\pi$/2) and ($\pi$,0). 
In the undoped compounds, both La$_2$CuO$_4$ and Ca$_2$CuO$_2$Cl$_2$, 
the broad peaks far below $E_{\rm F}$ represent the insulating gap 
with the $d$-wave-like dispersion~\cite{Ronning}. 
In the case of La$_{2-x}$Sr$_x$CuO$_4$ [Fig.~\ref{fig4}(b)], 
the features of the insulator remain at the same energy 
upon doping while new states are created inside the gap 
as seen in the ($\pi$,0) spectrum~\cite{Ino97,Ino00}. 
This duality indicates that the chemical potential is {\it fixed} upon doping. 
As summarized in Fig.~\ref{fig4}(a), the situation in the doped oxychloride 
is strikingly different. The evolution of the electronic states 
in this system can be understood roughly by a shift of the chemical potential 
to the top of the valence band. Intimately linked with the difference 
in the electronic evolution, the distribution of the spectral-weight 
in momentum space contrasts substantially between the two systems. 
As illustrated in Fig.~\ref{fig4}, the predominance of the low energy 
spectral weight in metallic La$_{1.9}$Sr$_{0.1}$CuO$_4$ 
occurs near ($\pi$,0) and (0,$\pi$)~\cite{Zhou}, while it occurs 
near ($\pi$/2,$\pi$/2) in the metallic Ca$_{1.9}$Na$_{0.1}$CuO$_2$Cl$_2$.

The duality of La$_{2-x}$Sr$_x$CuO$_4$, together with the 1D nature 
of the ($\pi$,0) Fermi surface 
and the lack of the chemical potential shift, 
has been linked with stripe formation~\cite{Ino97,Ino00,Zhou}; 
a phase separation into 1D charge and spin stripes. 
The rigid band shift of the oxychlorides suggests that antiferromagnetic 
fluctuations, originating from the parent insulator, 
instead dominate the metallic region for oxychlorides. 
Why is the oxychloride system so distinct from La$_{2-x}$Sr$_x$CuO$_4$? 
Structurally, the main difference between the two lies in the apical site 
(Cl vs. O) and the distortion of the CuO$_2$ planes 
(undistorted vs. distorted). These slight structural differences might help 
stabilize various different phases, 
resulting in enhanced antiferromagnetic correlations in the oxychlorides. 
In any event, the doped oxychloride system provides a new paradigm 
in the electronic evolution not previously observed in the cuprates. 
Surprisingly, however, the appearance of HTS upon doping 
does not crucially depend on which path leads 
the parent insulator to a metal. 

We would like to thank A. Ino, A. Fujimori, T. Tohyama, S. Maekawa, 
N. Nagaosa, S. Uchida, and K. Kitazawa for fruitful discussions. 
This work 
was partly supported by a Grant-in-Aid for Scientific Research from 
the MEXT, Japan. 
ARPES experiments 
were carried out at SSRL, a national user 
facility operated by Stanford University on behalf of the U.S. DOE. 

%------------------------------------
% references
%------------------------------------


\begin{references}
\vspace*{-5mm}
\bibitem{Ino97} 
 A. Ino {\it et al.}, 
 Phys. Rev. Lett. {\bf 79}, 2101 (1997). 
\bibitem{Ino00} 
 A. Ino {\it et al.}, 
 Phys. Rev. B {\bf 62}, 4137 (2000). 
\bibitem{Wells} 
 B. O. Wells {\it et al.},
 Phys. Rev. Lett. {\bf 74}, 964 (1995). 
\bibitem{LaRosa} 
 S. LaRosa {\it et al.}, 
 Phys. Rev. B {\bf 56}, R525 (1997).
\bibitem{Kim} 
 C. Kim {\it et al.}, 
 Phys. Rev. Lett. {\bf 80}, 4245 (1998). 
\bibitem{Ronning} 
 F. Ronning {\it et al.}, 
 Science {\bf 282}, 2067 (1998). 
\bibitem{Takagi} 
 H. Takagi, 
 Physica (Amsterdam) {\bf 341C-348C}, 3 (2000).
\bibitem{Hiroi94} 
 Z. Hiroi, N. Kobayashi, and M. Takano, 
 Nature (London) {\bf 371}, 139 (1994). 
\bibitem{Hiroi96} 
 Z. Hiroi, N. Kobayashi, and M. Takano, 
 Physica (Amsterdam) {\bf 266C}, 191 (1996). 
\bibitem{Kohsaka}
 Y. Kohsaka {\it et al.} 
 J. Am. Chem. Soc., in press. 
\bibitem{Aebi}
 P. Aebi {\it et al.} 
 Phys. Rev. Lett. {\bf 72}, 2757 (1994). 
\bibitem{Dessau}
 D. S. Dessau {\it et al.}, 
 Phys. Rev. Lett. {\bf 66}, 2160 (1991). 
\bibitem{Feng}
 D. L. Feng {\it et al.}, 
 Science {\bf 289}, 277 (2000). 
\bibitem{Ding}
 H. Ding {\it et al.}, 
 Phys. Rev. Lett. {\bf 76}, 1533 (1996). 
\bibitem{Campuzano}
 J. C. Campuzano {\it et al.}, 
 Phys. Rev. Lett. {\bf 83}, 3709 (1999). 
\bibitem{Lanzara}
 A. Lanzara {\it et al.},
 Nature (London) {\bf 412}, 510 (2001). 
\bibitem{Marshall} 
 D. S. Marshall {\it et al.}, 
 Phys. Rev. Lett. {\bf 76}, 4841 (1996). 
\bibitem{Furukawa} 
 N. Furukawa, T. M. Rice, and M. Salmhofer, 
 Phys. Rev. Lett. {\bf 81}, 3195 (1998). 
\bibitem{Wen} 
 X. -G. Wen and P. A. Lee, 
 Phys. Rev. Lett. {\bf 76}, 503 (1996). 
\bibitem{Chakravarty} 
 S. Chakravarty {\it et al.}, 
 Phys. Rev. B {\bf 63}, 094503 (2001).
\bibitem{Zhou} 
 X. J. Zhou {\it et al.}, 
 Science {\bf 286}, 268 (1999). 
\end{references}
\end{document}